# Defect Control via Cu Enrichment Enhances Multifunctional Properties in the Polar Semiconductor $Cu_{1+x}Mn_{1-y}SiTe_3$

Subrata Ghosh[1,2,+,*], Yu Liu[1,2,+], Saugata Sarker[3], Boyang Zheng[1,2], Sreekant Anil[3], Soumi Mondal[2], Yuxi Zhang[3], Sai Venkata Gayathri Ayyagari[3], Mingyu Xu[4], Yingdong Guan[1,2], Tsung-Han Yang[5], Xiaoping Wang[5], Vincent H. Crespi[1,2], Nasim Alem[3], Weiwei Xie[4], Venkatraman Gopalan[2,3], Qiang Zhang[5*], and Zhiqiang Mao[1,2,3,*]

*[1]2D crystal consortium, Materials Research Institute, Pennsylvania State University, University Park, PA 16802, USA.*
*[2]Department of Physics, Pennsylvania State University, University Park, PA 16802, USA.*
*[3]Department of Materials Science and Engineering, Pennsylvania State University, University Park, PA 16802, USA*
*[4]Department of Chemistry, Michigan State University, East Lansing, MI 48824, USA.*
*[5]Neutron Scattering Division, Oak Ridge National Laboratory, Oak Ridge, TN 37831, USA.*

*Corresponding authors' email*: smg7204@psu.edu (S.G.); zhangq6@ornl.gov (Q.Z.); zim1@psu.edu (Z.M.)

**Abstract**

Polar materials have recently attracted significant interest due to their rich multifunctional properties. The chalcogenide polar semiconductor $Cu_{1-x}Mn_{1+y}SiTe_3$ (Cu-deficient, $0.04 \leq x \leq 0.26$; $0.03 \leq y \leq 0.15$) is an emerging multiferroic system in which electric polarization is coupled to magnetization. However, its macroscopic ferroelectric polarization is strongly suppressed due to the presence of a high density of stacking faults. In this work, we demonstrate that these crystal defects, likely originating from non-stoichiometry, can be substantially reduced by increasing the Cu content. Cu-enriched samples, $Cu_{1+x}Mn_{1-y}SiTe_3$ ($0.04 \leq x \leq 0.3$; $0.13 \leq y \leq 0.31$), crystallize in a noncentrosymmetric monoclinic structure (space group *Pm*) as the Cu-deficient counterpart but show a nearly stacking-fault-free phase, which is attributed to the emergence of an interstitial site. Consequently, the Cu-enriched samples show a pronounced enhancement of the second-harmonic generation (SHG) response compared to Cu-deficient compositions. Magnetically, the Cu-enriched crystals retain long-range antiferromagnetic order with a Néel temperature of $T_N \sim 33$ K without a glassy state but manifest a distinct spin-flop transition along the polar *b*-axis that is absent in the Cu-deficient compositions. Furthermore, the electronic ground state evolves from insulating to doped semiconducting behavior upon Cu enrichment. Together, these results establish this material system as a unique and versatile platform for elucidating the interplay among composition, crystal defects, and multifunctional properties, offering a route to design magnetic polar systems with tunable quantum functionalities.

[+]S.G. and Y. L. contributed equally to this work.

## Introduction

Incorporating multiple physical phenomena, particularly those that are mutually exclusive, within a single-phase material represents a compelling strategy for realizing emergent functionalities [1,2]. For instance, the coexistence of ferroelectricity and metallicity is conflicting in a single-phase material due to the screening effect of itinerant charge carriers on long-range dipole-dipole interactions. However, Anderson and Blount first proposed that ferroelectricity can persist in a metal based on symmetry considerations, leading to the emergence of polar metals [3]. This discovery has stimulated extensive efforts to identify and design polar materials, motivated by their novel electronic states and potential quantum functionalities [1,2,11–16,3–10]. Several strategies have been employed to design the polar metals, including the breaking of inversion symmetry through displacements of atoms whose electronic degrees of freedom are decoupled from the states at the Fermi level [11,12,14], introducing metallicity into polar structures via carrier doping and designing artificial layered structures [1,2]. Furthermore, polar materials with coexisting magnetic ordering have enabled the discovery of various multifunctional properties, such as multiferroicity with strong magneto-electric (ME) coupling, which have potential applications in electric field control of magnetization or vice versa [13,17].

To date, extensive research has explored the exotic multifunctional properties of polar materials. For instance, the transition metal-based dichalcogenide (TMD), $WTe_2$, is the first known ferroelectric metal to demonstrate switchable spontaneous out-of-plane electric polarization under an external electric field in bi- and tri-layered [4] as well as bulk [15] forms. It also retains the ferroelectric order down to the two-dimensional (2D) limit [4], prompting the search for layered multiferroics among TMDs and halides [18–20]. Moreover, Sakai *et al.* [10] observed a significant enhancement of thermopower close to the critical region between the polar and nonpolar metallic phases in a polar semimetal 1T′-$MoTe_2$ with a tunable polar transition. Rischau et al. demonstrated the unconventional superconductivity in a distortive polar metal $Sr_{1-x}Ca_xTiO_{3-\delta}$ [16], where ferroelectric order coexists with superconductivity within a specific doping/carrier density regime. Polar materials are proposed to host nontrivial topological properties as well [6,7]. Additionally, the interplay between magnetic and polar order in a magnetic polar material opens an avenue for realizing the exotic phases [5,21–23]. For example, Zhang *et al.* [5] reported the coexistence of intrinsic ferromagnetism, polar distortion, and metallicity in a quasi-2D system, $Ca_3Co_3O_8$. Consequently, these accelerated discoveries underscore the crucial need to identify new polar materials with novel multifunctional properties.

We recently reported the chalcogenide semiconductor, $Cu_{1-x}Mn_{1+y}SiTe_3$ ($0.04 \leq x \leq 0.26$; $0.03 \leq y \leq 0.15$) (termed as Cu-deficient), as a promising multiferroic material with a strong ME coupling [9]. However, the Cu-deficient systems show high-density stacking faults, likely originating from their non-stoichiometric composition, which suppresses the electrical polarization. In this work, we show that the stacking faults of this material are controlled by its crystal composition and can be significantly reduced by increasing the Cu concentration. We find that although the Cu-rich samples, $Cu_{1+x}Mn_{1-y}SiTe_3$ ($0.04 \leq x \leq 0.3$; $0.13 \leq y \leq 0.31$), adopt a monoclinic polar structure with a noncentrosymmetric space group of *Pm,* similar to the Cu-deficient system, their structure is characterized by an emerging interstitial atomic site that leads to a the substantial suppression of stacking faults across the phase matrix. Consequently, the SHG polarimetry data can be well described by a single-domain model with monoclinic m point-group symmetry. This result confirms the improved crystal quality, in clear contrast to the Cu-deficient sample, for which a reliable single-domain fit was not obtained [9]. Furthermore, the SHG data fit well with a monoclinic 'm' point-group symmetry using a single-domain model, explicitly verifying the significant improvement of crystal quality.

Beyond the structural properties, Cu-enrichment profoundly impacts the magnetic and electronic properties of the material. Cu-enriched crystals show long-range antiferromagnetic ordering below 33 K, accompanied by a distinct spin-flop transition along the polar *b*-axis. Notably, while Cu deficient crystals exhibit a signature of short-range magnetic ordering and a glassy state, these features are absent in the Cu-enriched crystals. While the bandgap of Cu-enriched crystal remains similar to that of Cu-deficient crystal, it shows a metal-like characteristic with an enhanced carrier density upon Cu doping. This results in high leakage current, which limits the direct measurement of electrical polarization and magneto-electric coupled parameters. However, the magnetoresistance ($MR_{xx}$) shows a noticeable cusp-like feature at low temperatures, attributed to the weak antilocalization (WAL), which manifests the presence of large spin-orbit coupling that can potentially increase the magnetoelectric coupling of the material. These results advance our understanding of the complex interplay between crystal composition, crystal defects, and functional properties and provide a route for designing functional polar materials with non-stoichiometric compositions.

# Results and Discussion

## A. Crystal structure, microstructure, and electronic properties of $Cu_{1+x}Mn_{1-y}SiTe_3$

We first identify the distinct crystal structure of the Cu-enriched crystal in comparison to the Cu-deficient system. Room-temperature single-crystal X-ray and neutron diffraction measurements reveal that Cu-enriched samples crystallize in a monoclinic, noncentrosymmetric structure with space group *Pm* (No. 6), as shown in **Fig. 1A**. Similar to the Cu-deficient counterparts (**Fig. 1B**), the crystal structure features ethane-like $Si_2Te_6$ units consisting of Si-Si dimers coordinated by Te atoms. However, in the Cu-enriched phase, the coordination environments of Cu and Mn differ from those in the Cu-deficient samples. While in Cu-deficient samples, Mn coordination is octahedral and Cu coordination is tetrahedral, in the case of Cu-enriched samples, both Cu and Mn adopt distorted octahedral and tetrahedral geometries. Moreover, both coordination polyhedra display substantial distortion, atomic-site mixing, and partial vacancies. The incorporation of excess Cu leads to partial occupation of mixed atomic sites and induces Mn displacement toward interstitial positions, resulting in the emergence of additional atomic sites within the framework. The structural parameters at room temperature are summarized in **Table S1**. This structural modification enhances the three-dimensional connectivity of the Cu-enriched phase, in contrast to the quasi-two-dimensional framework characteristic of the Cu-deficient analogues. The increased structural coherence also reduces stacking faults commonly observed in the Cu-deficient samples, as discussed below.

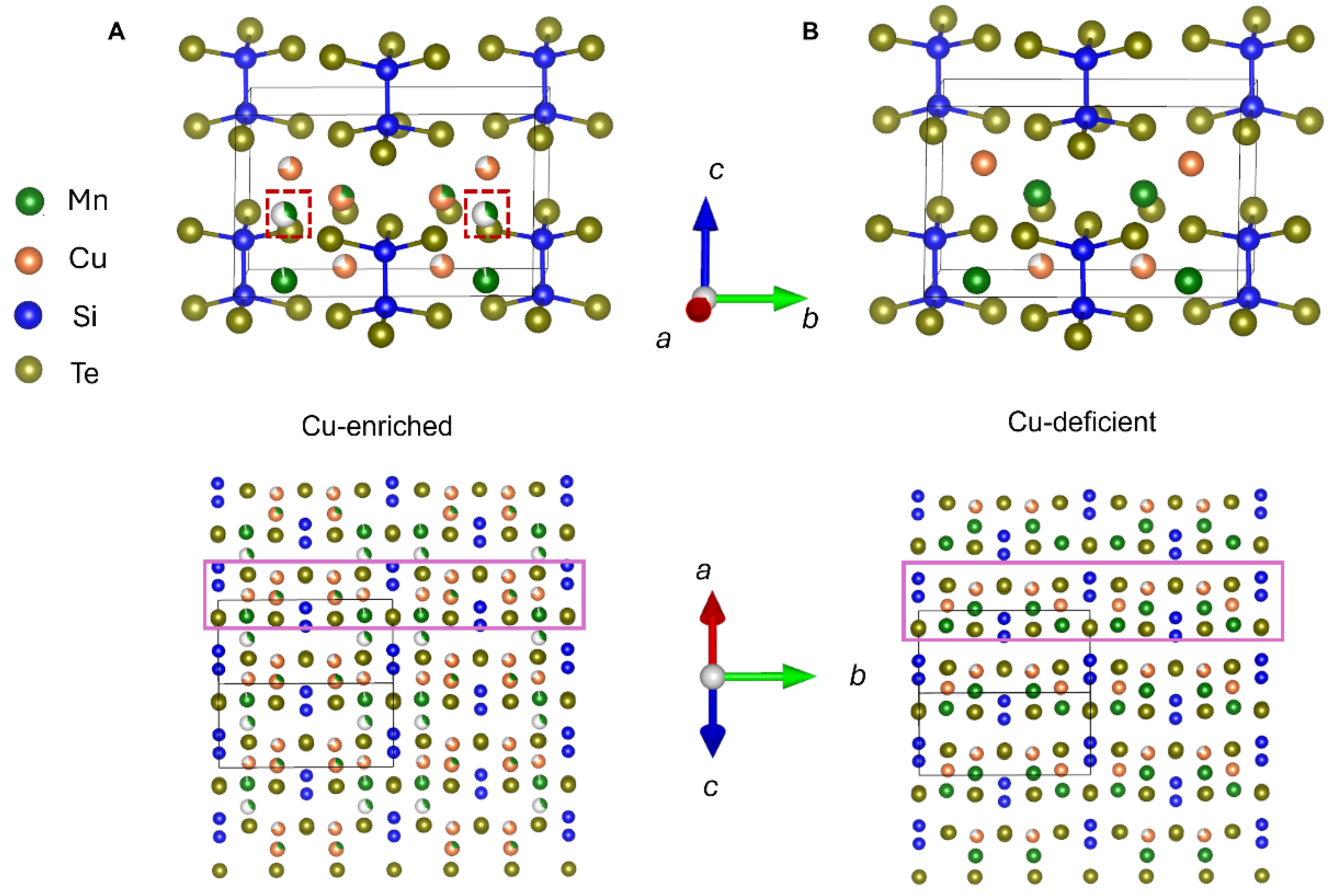

**Figure 1: Crystal structure from single-crystal X-ray diffraction.** Crystal structure of (A) Cu-enriched (Left panel) and (B) Cu-deficient (right panel) from single-crystal X-ray refinement. The coordination environments of Cu and Mn in the Cu-enriched sample are different from those in the Cu-deficient sample. Cu and Mn atoms adopt distorted octahedral and tetrahedral geometry. The occupation of excess Cu leads to the emergence of additional atomic sites within the framework.

The microstructural characterization of the Cu-enriched crystals, $Cu_{1.14}Mn_{0.77}SiTe_{2.89}$ (S#1), reveals a substantial suppression of crystallographic defects, particularly stacking faults, upon Cu doping. The selected-area electron diffraction (SAED) pattern along the [010] zone axis (Figure 2A) shows sharp diffraction spots without any streaks, indicating that stacking faults are significantly suppressed compared to the Cu-deficient system [9]. This is further confirmed from the high-resolution high-angle annular dark field (HAADF) images from scanning transmission electron microscopy (STEM), which show the absence of stacking faults in the phase matrix (**Fig. 2B**), a feature that was prominently observed in the Cu-deficient sample [9]. The Te-sublattice arrangement in the HAADF-STEM image matches the single-crystal refined crystal structure. The incorporation of excess Cu atoms renders the Cu-rich phase more three-dimensional and bulk-like, rather than strongly layered. As stacking faults are commonly associated with layered structures, the increased structural connectivity in the Cu-enriched crystals suppresses the formation of stacking faults. The suppression of stacking faults in Cu-enriched crystals is attributed to the emergence of an interstitial atomic site induced by Cu doping, which enhances structural coherence. Such defect mitigation clearly demonstrates that Cu doping significantly improves the crystalline quality of the material. Consequently, we observed an enhanced optical SHG response as discussed below.

The observed optical SHG response in Cu-enriched crystals further confirms the retention of a noncentrosymmetric crystal structure. As shown in **Figure 2C,** the SHG polar plot of sample (S#1) can be well fitted using a monoclinic '*m*' point-group symmetry within a single-domain approximation. The SHG power-dependent measurements (**Fig. 2D**) demonstrate that the Cu-enriched crystal exhibits a significantly stronger SHG response, which is around an order larger than those of the nearly stoichiometric and Cu-deficient variants. Furthermore, as the SHG intensity is proportional to the magnitude of second-order nonlinear polarization of the material (**see Note S1)**, a significant increase in SHG output reflects an enhancement of induced nonlinear polarization. Additionally, the enhanced SHG response also indicates improved crystal quality and substantial suppression of defects in Cu-enriched crystals, consistent with microstructural observations. The correlation between suppressed stacking faults and the enhanced SHG response suggests that crystallographic defects play a crucial role in limiting the ferroelectric polarization in the Cu-deficient system.

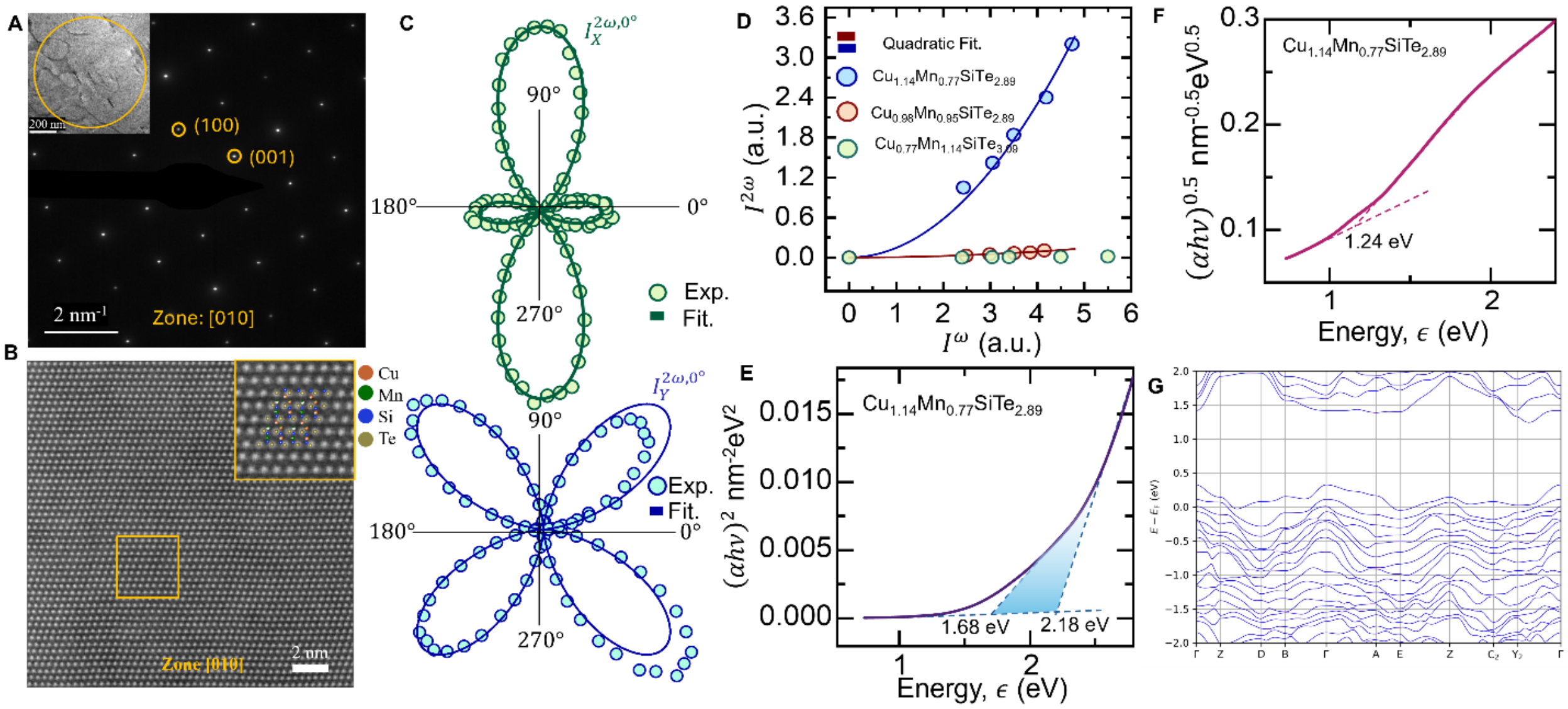

**Figure 2: SHG, electron diffraction pattern, and ellipsometry measurement of $Cu_{1.14}Mn_{0.77}SiTe_{2.89}$ (S#1)**. (A) Selected area electron diffraction (SAED) pattern of S#1 along (010) zone axis. (B) STEM image along the [010] zone axis clearly showing the structure matching with the X-ray refined structure, and the crystal is free from excessive stacking faults. (C) SHG polarimetry of the Cu-enriched sample as a function of incident polarization, measured under two different analyzer configurations. (D) Intensity of the second harmonic as a function of fundamental intensity for the Cu-enriched sample. (E–F) Tauc plot analysis of the Cu-enriched sample, used to extract both the direct and indirect band gaps. (G) The band structure of the relaxed model ($Cu_6Mn_2Si_4Te_{12}$), where the major elements are used for partial occupancy sites. The band structure is spin-degenerate. The direct/indirect band gaps are 1.09/0.92 eV.

Spectroscopic ellipsometry measurements confirm the semiconductor nature of the Cu-enriched sample. Analysis of the Tauc plot, as shown in **Figs. 2E** and **2F,** yields an indirect bandgap of ~1.24 eV and a direct bandgap in the range of ~1.68-2.18 eV. To further elucidate the electronic structure of the Cu-enriched samples, we performed density functional theory (DFT) calculations (**see Note S2**). The calculated band structure of the relaxed Cu-enriched model (**Fig. 2G**, for the unrelaxed model see **Fig. S1**) reveals direct/indirect gaps 1.09/0.92 eV, which closely resemble those obtained for the Cu-deficient model (direct/indirect gaps 1.10/0.94 eV) [9]. Notably, the Fermi level in the Cu-enriched model lies in the valence band, which indicates a *p*-doped indirect-gapped semiconducting feature, in agreement with experimental observations presented below. Although DFT calculation typically underestimates the bandgap values, both the calculated and measured optical bandgap of the Cu-enriched sample remain comparable to those of the Cu-deficient system [9]. This suggests that Cu doping primarily modifies the defect landscape and carrier concentration while preserving the semiconducting electronic structure.

## B. Magnetic properties of $Cu_{1+x}Mn_{1-y}SiTe_3$

To elucidate the effect of Cu doping on magnetic properties, we performed *dc* magnetization and specific heat capacity measurements of the Cu-enriched system (S#2: $Cu_{1.10}Mn_{0.80}SiTe_{2.80}$). The Cu-deficient crystals demonstrated short-range magnetic correlations below 60 K, followed by the establishment of a long-range antiferromagnetic (AFM) order at $T_N$ ~ 35 K. Below ~15 K, a weak ferromagnetic component originated from a canted AFM state with the net moment lying in the *ab* plane [9]. **Figures 3(A-B)** show the temperature-dependent magnetization (*M-T*) curves of the Cu-enriched crystal, measured under zero-field cooled (ZFC) and field-cooled (FC) conditions in a 1 kOe magnetic field along the *a*-axis, *b*-axis, and *c*-axis. The material undergoes AFM ordering transition at $T_N$ ~33 K, marked by a sharp cusp in the dc magnetization curve (**Fig. 3B**). The magnetic susceptibility in the high-temperature regime (100 K-300 K) follows the Curie-Weiss law (**Fig. S2**), yielding a Curie-Weiss temperature ($T_{CW}$) ~ -121 K, consistent with AFM behavior below $T_N$. The effective magnetic moment is determined to be $\mu_{eff}$ ~ 5.32$\mu_B$, close to the expected localized Mn moment. Moreover, the frustration index, $f$ (= $T_{CW}/T_N$), in the Cu-enriched sample is low (~3.67), suggesting a weakly frustrated magnetic lattice [24], in contrast to moderate magnetic frustration and glassy behavior, observed in the Cu-deficient sample [9]. Magnetic relaxation measurements further confirm the absence of a spin-glass state (**see Note S3** and **Fig. S3)**. In addition, the zero-field specific heat capacity ($C_p$) shows a clear anomaly at 33 K, confirming the emergence of long-range AFM order below 33 K (**Fig. 3C**). Together, these results demonstrate that Cu-doping preserves long-range AFM order while suppressing short-range magnetic correlations and glassy behavior, likely due to the significantly reduced stacking faults.

The anisotropic field-dependent magnetization further reveals the complex magnetic behavior of the Cu-enriched crystal. **Figures 3(D-E)** show the isothermal magnetization (*M* vs. *H*) curves measured at 5 K with the magnetic field applied along the *a*-axis, *c*-axis, and *b*-axis. For *H*

// *a* and *H* // *c*, *M-H* curves show typical colinear AFM behavior (**Fig. 3D**). In contrast, for *H* // *b*, a sudden jump in magnetization ($\Delta M$) near 3.5 T is observed, indicating a spin-flop transition from an AFM state to a canted AFM state (**Fig. 3E**). This feature is typically observed due to the rotation of canted spin in the external field. The metamagnetic phase transition is observed till $T_N$, and beyond that it shows usual paramagnetic behavior (**Fig. 3F**). The spin-flop field ($H_{FC}$) is observed to shift toward the higher magnetic field value with increasing temperature (inset of **Fig. 3F)**. The field-induced metamagnetic transition is further evidenced by the sign reversal of magnetic entropy change ($\Delta S_m$) at $H_{FC}$ for *H* // *b* (see **Note S4**, **Fig. S4,** and **S5**). Moreover, the anisotropic magnetic behavior can be understood in terms of canted AFM with the net magnetic moment aligned along the *c* axis.

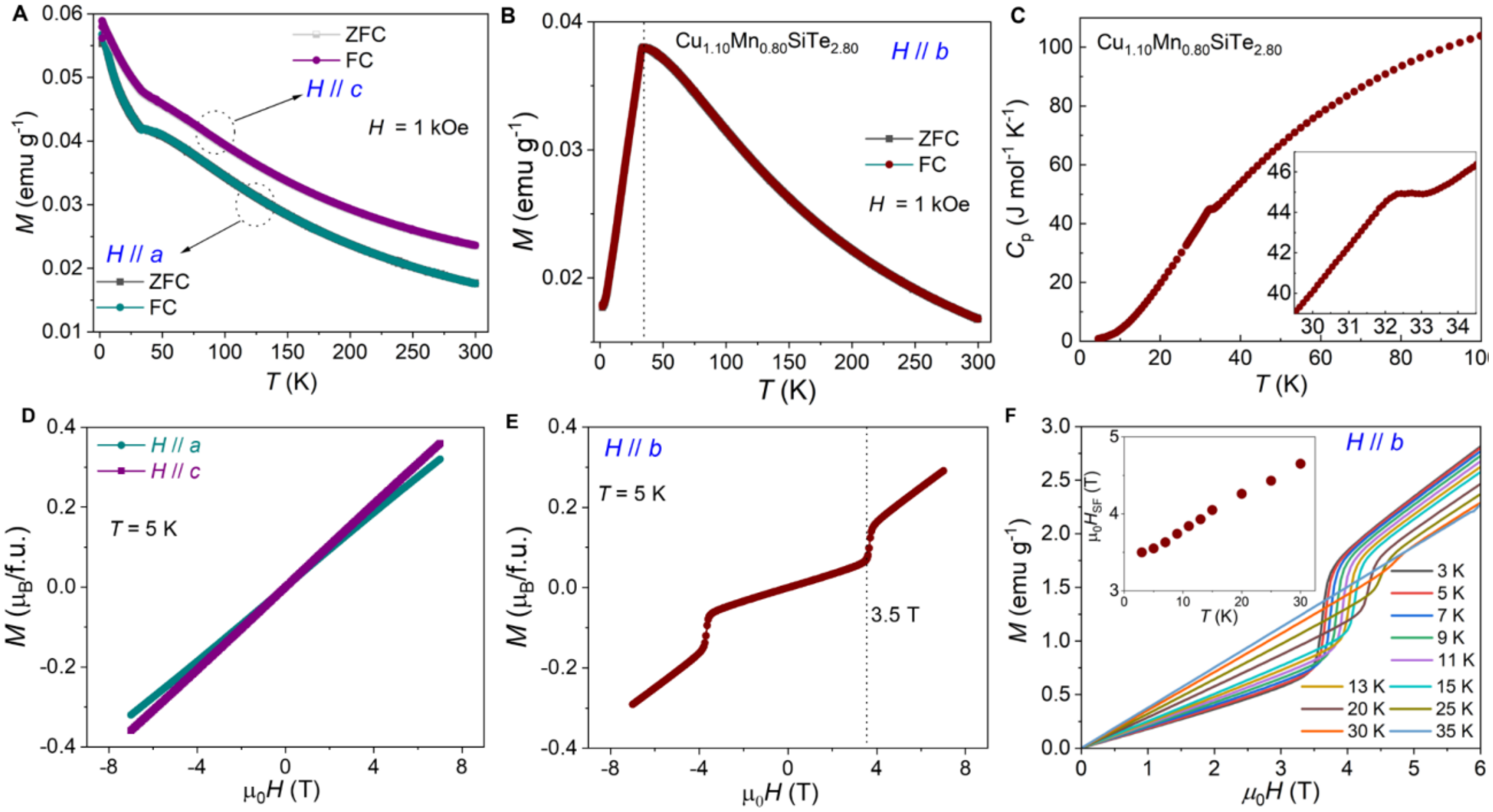

**Figure 3: Magnetic properties and specific heat capacity of $Cu_{1.10}Mn_{0.80}SiTe_{2.80}$ (S#2)**: Temperature dependence of dc magnetization in the presence of 1kOe magnetic field under ZFC and FC conditions, with the applied field parallel to (A) *a*-axis, *c*-axis, and (B) polar *b*-axis. (C) Specific heat capacity ($C_p$) as a function of temperature under zero field. [*Inset*: the zoomed image showing the anomaly]. Isothermal *M-H* curves at 5 K with the field applied (D) *H* // *a*, and *H* // *c* axis, and (E) *H* // *b* axis. (F) Isothermal *M-H* curves at different temperatures under *H* // *b* condition [*Inset*: The shifting of spin-flop field towards the higher field with increasing temperature]

To resolve the magnetic structure, single-crystal neutron diffraction experiments were carried out at 100 K ($T > T_N$) and 5 K ($T < T_N$). At 5 K, although the crystal structure remains in the monoclinic *Pm* space group with no detectable structural transition, distinct magnetic reflections emerge. **Figure 4 (A-B)** displays the single-crystal neutron diffraction patterns in the representative (HK0) plane and (0KL) plane at 100 K ($T > T_N$) and 5 K ($T < T_N$), respectively. The magnetic reflections, highlighted by circles and identified by the enhanced intensity arising from magnetic contributions, are clearly visible at 5 K. The temperature dependence of the intensity for the low-Q magnetic reflections 024 and 025 in **Fig. 4D** indicates the magnetic transition temperature around $T_N = 33$ K, in agreement with magnetization and heat capacity measurements. Moreover, when comparing the data at 100 K with that at 5 K, no appreciable broadening of the magnetic reflection linewidth is observed at 5 K, indicating that long-range magnetic order is established below $T_N$.

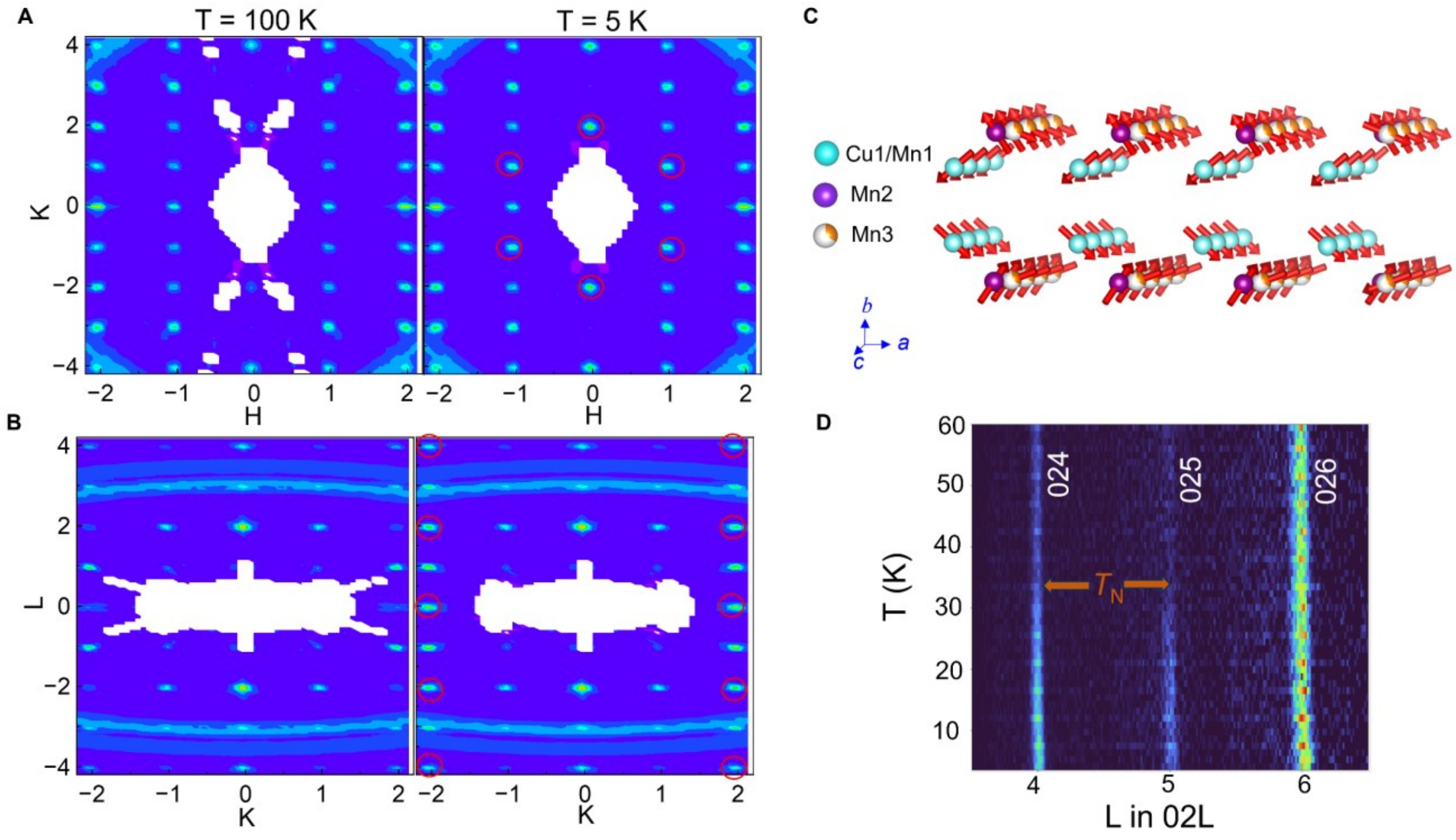

**Fig. 4. Magnetic structure of $Cu_{1.14}Mn_{0.78}SiTe_3$ (S#3) determined by single-crystal neutron diffraction.** (**A-B**) Comparison of single-crystal neutron diffraction patterns in the (HK0) plane and (0KL) plane, respectively, at 100 K ($T > T_N$) and 5 K ($T < T_N$). The circles highlight strong magnetic Bragg reflections. (**C**) Magnetic structure at 5 K (only magnetic atoms displayed). (**D**) Temperature dependence of the intensity of nuclear and/or magnetic Bragg reflections 024, 025, and 026.

All the magnetic reflections can be indexed by the crystallographic unit cell, yielding a magnetic propagation vector $\boldsymbol{k} = 0$. The determined magnetic structure is illustrated in **Fig. 4C**. There is no evidence of the ordered moments on Cu1 or Cu2 sites. Ordered moments were found at the mixed Mn1/Cu1 site, the fully occupied Mn2 site, and the partially occupied Mn3 site. The spins at these three magnetic sites must be canted to achieve a good fit to all the 14200 nuclear and magnetic reflections (including redundant reflections). The AFM order belongs to the magnetic space group *Pm* (BNS No. 6.18), without a net ferromagnetic component. For each of these three magnetic sites, the projections onto the *ac* plane exhibit antiferromagnetic alignment along the *b*-axis while aligning ferromagnetically along the *a*- and *c*- axes. This interprets the anisotropic magnetization results with the field parallel to these three crystallographic axes presented in **Fig. 3 (A–B)**. The magnetic moments at Mn1/Cu1, Mn2, and Mn3 sites are determined to be (2.0, -1.7, 1.1) $\mu_B$, (-1.24, 1.96, 0.86) $\mu_B$, and (2.5, -1, 1.3) $\mu_B$, respectively, leading to an ordered moment of 2.6(3) $\mu_B$. This magnetic structure in $Cu_{1.14}Mn_{0.78}SiTe_3$ is markedly different from a collinear AFM order reported for the Cu-deficient sample [9] due to the additional magnetic sites, site disorder, and site deficiencies in the Cu-rich sample.

The substantially suppressed stacking faults with the enhanced SHG response and robust long-range AFM ordering indicate the overall improvement of structural coherence and magnetic properties in the Cu-enriched system. Such enhanced crystalline order may promote stronger electric polarization and magnetoelectric coupling. However, Cu doping induces a high carrier density in the system, driving the system toward a doped semiconductor with weak metal-like behavior. The resulting high leakage current complicates the measurement of intrinsic magnetoelectric properties. Therefore, understanding the electrical transport properties of Cu-enriched samples is essential and is discussed next.

## C. Magneto-transport properties of $Cu_{1+x}Mn_{1-y}SiTe_3$

We investigated the electrical transport properties of Cu-enriched samples, $Cu_{1.16}Mn_{0.73}SiTe_{2.91}$ (S#4) and $Cu_{1.29}Mn_{0.77}SiTe_{2.99}$ (S#5). In contrast to the Cu-deficient sample, which demonstrates an insulating electronic ground state, strongly coupled to its magnetic ordering [9], the Cu-enriched samples show a doped semiconducting behavior. Notably, no noticeable anomaly in the electrical resistivity is observed across $T_N$ (**Fig. S6A** and see also **Fig. S10** of ref. [9]). **Figure 5A** shows the magnetic field dependence of in-plane magnetoresistance ($MR_{xx}$) for S#4 (see also **Fig. S6B** for S#5**),** defined as $MR_{xx} = [\rho_{xx}(H) - \rho_{xx}(0)]/\rho_{xx}(0)$ where $\rho_{xx}(H)$ and $\rho_{xx}(0)$ are the longitudinal resistivity in the presence and absence of an applied magnetic field, respectively. All the $MR_{xx}$ data are symmetrized to eliminate the contribution of Hall resistivity to it. The Cu-enriched sample exhibits a small and non-saturating quasilinear $MR_{xx}$ (~ 0.44% at 3K, 9T). The relatively weak $MR_{xx}$ response suggests low carrier mobility in Cu-enriched samples.

Interestingly, we observed a cusp-like behavior in $MR_{xx}$ that gradually weakens with increasing temperature, attributed to the quantum interference phenomenon, i.e. weak antilocalization (WAL). To elucidate the WAL phenomenon, magnetoconductivity is calculated using the expression, $\Delta\sigma_{xx}(B) = \sigma_{xx}(B) - \sigma_{xx}(0)$ where $\sigma_{xx}(B) = \rho_{xx}/(\rho_{xx}^2 + \rho_{xy}^2)$, $\sigma_{xx}(0)$ is the zero-field conductivity and $\rho_{xy}$ is the Hall resistivity (**Fig. 5B)**. The WAL is analyzed with the modified Hikami-Larkin-Nagaoka (HLN) model. Although the HLN model was derived for two-dimensional (2D) systems, it is frequently employed to describe WAL in 3D bulk materials [25,26]. In the modified HLN expression, a $\gamma B^2$ term related to parabolic conductance is introduced to incorporate spin-orbit scattering and elastic scattering that leads to [27,28]

$$\Delta\sigma_{xx}(B) = \frac{-\alpha e^2}{2\pi^2\hbar}\left[\psi\left(\frac{1}{2} + \frac{B_\varphi}{B}\right) - ln\left(\frac{B_\varphi}{B}\right)\right] - \gamma B^2 \quad (1)$$

where $\psi\ (x)$ is the digamma function and $B_{\varphi} = \frac{\hbar}{4eL_{\varphi}^2}$ is the characteristic field, and $L_{\varphi}$ is the phase coherence length. As shown in **Fig. 5B**, the modified HLN model provides an excellent fit for the experimental data. The extracted $L_{\varphi}$ is plotted as a function of temperature in the inset of **Fig. 5C**. At 3K, $L_{\varphi}$ is ~31.7 nm and decreases monotonically with an increase in temperature as the enhanced inelastic scattering disrupts the phase coherence. The estimated $L_{\varphi}$ is significantly smaller than the crystal's thickness, implying the 3D nature of the WAL effect. Moreover, the trend of $L_{\varphi}$ with temperature can be explained by considering the electron-electron scattering (*e-e*) and electron-phonon (*e-ph*) scattering, which follows the following relationship,

$$\frac{1}{L_{\varphi}^2(T)} = \frac{1}{L_{\varphi}^2(0)} + A_{e-e}T + A_{e-ph}T^2 \tag{2}$$

where $L_{\varphi}(0)$ is the zero-temperature phase coherence length. $A_{e-e}$ and $A_{e-ph}$ represent *e-e* and *e-ph* scattering coefficients, respectively. **Figure 5(C)** shows the well fit of the data with estimated $L_{\varphi}(0)$ = 43.3 nm, $A_{e-e}$ ~$1.16 \times 10^{-4}$ nm$^{-2}$ K$^{-1}$ and $A_{e-ph}$ ~$1.3 \times 10^{-6}$ nm$^{-2}$ K$^{-2}$. The larger value of $A_{e-e}$ suggests that *e-e* scattering dominates over the *e-ph* in the inelastic scattering process. The observed WAL effect is attributed to the strong spin–orbit coupling in this Cu-enriched system, which may also influence magnetoelectric coupling by providing a microscopic link among spin order, lattice distortions, and electric polarization.

Further, we also measured the magnetic field dependence of Hall resistivity ($\rho_{xy}$) at various temperatures (shown in **Fig. 5D** for S#4 and in **Fig. S6C** for S#5). The linear behavior of $\rho_{xy}$ with the magnetic field can be well described by a single-band transport model. The extracted Hall coefficient ($R_H$) remains positive throughout the measured temperature range up to 300 K, indicating hole-like bands dominate the electrical transport. The carrier density ($n_H$) and carrier mobility ($\mu$) of the material were determined using $n_H = 1/(eR_H)$ and $\mu = R_H/\rho_{xx}(0)$. For sample S#4, $n_H$ is ~$7 \times 10^{17}$ cm$^{-3}$, which remains nearly temperature independent up to 200 K and then gradually increases till room temperature to ~$1.1 \times 10^{18}$ cm$^{-3}$ (**Fig. 5E**). A similar trend is observed for S#5 (**Fig. S6D**), although with $n_H$ approximately one order of magnitude larger, demonstrating that carrier concertation can be effectively tuned via increasing Cu content. The Hall mobility is estimated to be relatively low ~1.7 cm$^2$V$^{-1}$s$^{-1}$, which accounts for the weak $MR_{xx}$ response in the Cu-enriched sample, as the magnetoconductivity for the single-band model can be approximated as, $\sigma(\mu_0 H) \sim n\mu/[1 + \mu^2(\mu_0 H)^2]$ [26,29,30].

To independently verify the nature of the dominant charge carriers, we performed thermopower measurements on $Cu_{1.24}Mn_{0.91}SiTe_{3.03}$ (S#6). Unlike Hall measurements, thermopower measurements do not require an applied magnetic field and therefore provide a more reliable probe of carrier type. The longitudinal Seebeck coefficient ($S_{xx}$) measured in the longitudinal direction (polar axis) shows a positive value throughout the measured temperature regime up to 300 K, confirming hole-dominated transport, consistent with the Hall effect measurement (**Fig. 5F**).

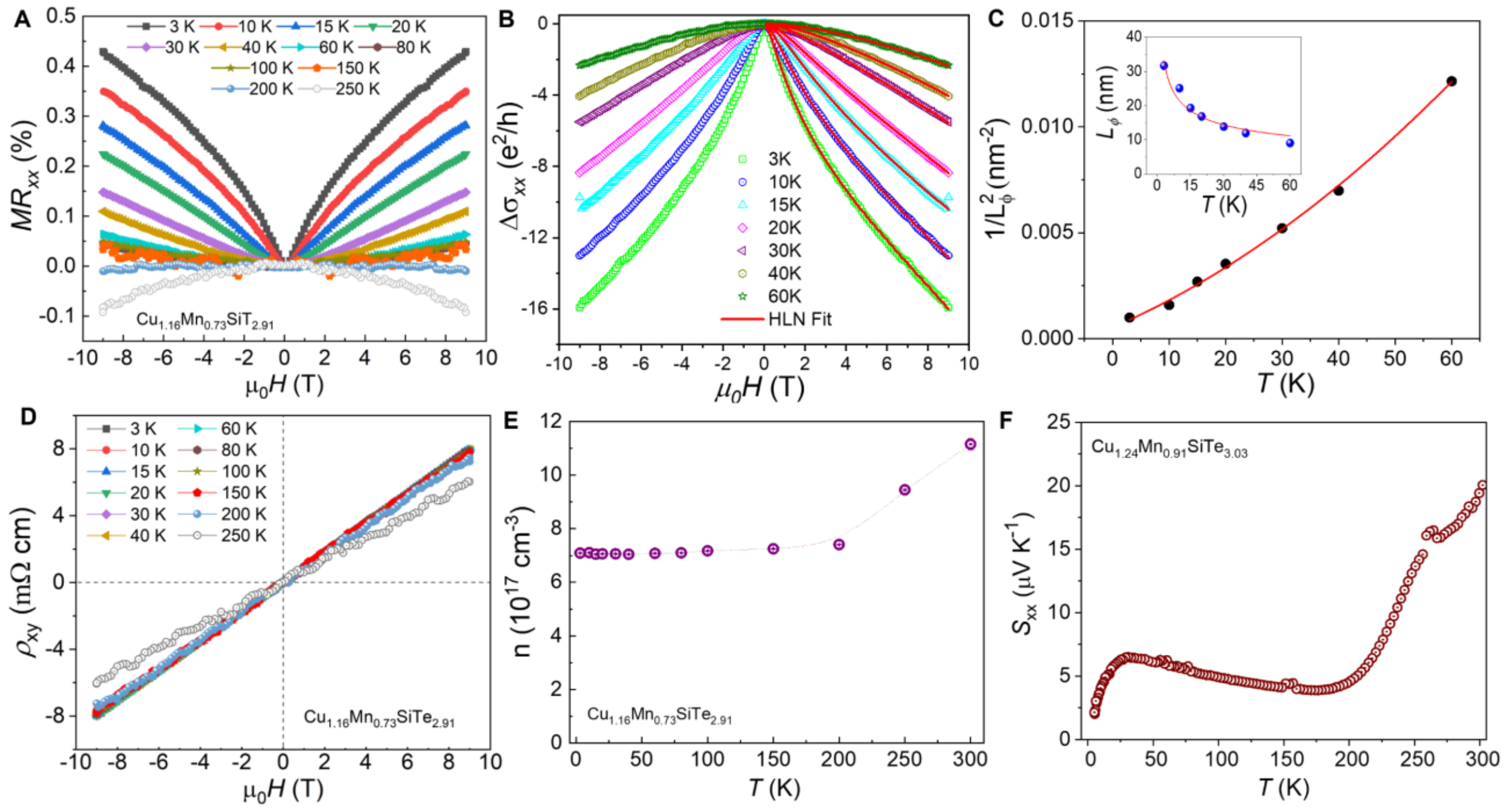

**Figure 5: Magnetotransport properties, and Seebeck coefficient of $Cu_{1+x}Mn_{1-y}SiTe_3$**: The magnetic field dependence of in-plane (A) magnetoresistance and (B) magnetoconductivity of $Cu_{1.16}Mn_{0.73}SiTe_{2.91}$ (S#4) at various temperatures. The magnetic field is applied along the out-of-plane direction of the sample. (C) The estimated phase coherence length from HLN fits (Eq. 1) as a function of temperature is plotted in the inset and $1/L_\varphi^2(T)$ vs. $T$ data is well fitted with Eq. 2. (D) The Hall resistivity of S#4 sample at various temperatures. (E) The calculated carrier density as a function of temperature for the S#4 sample. (F) Longitudinal Seebeck coefficient as a function of temperature, measured along the polar axis, of $Cu_{1.24}Mn_{0.91}SiTe_{3.03}$ (S#6) sample.

## CONCLUSION

In summary, the $Cu_{1+x}Mn_{1-y}SiTe_3$ system preserves the key characteristics of the polar $Cu_{1-x}Mn_{1+y}SiTe_3$ multiferroic family while the optimization of chemical compositions appears to suppress the stacking faults within the phase matrix. This results in enhanced structural coherence and a pronounced increase in optical second harmonic generation response, indicative of enhanced polarization. While long-range antiferromagnetic ordering is preserved in the Cu-enriched system, the emergence of a field-induced spin flop transition underscores the feasibility of accessing distinct magnetic phases in a polar lattice via tuning chemical composition. The excess Cu in the system introduces itinerant hole carriers, driving the material toward a doped semiconductor with weak metal-like transport behavior. Although the elevated electrical conductivity introduces leakage current challenges for conventional ferroelectric and magnetoelectric characterization in bulk single crystals, thinning down the sample to the nanometer scale can possibly facilitate the probing of ferroelectricity in this metal-like sample. Furthermore, it can also be anticipated that the coexistence of such a magnetic phase in a polar material may provide new avenues for exploring multiferroic behavior and emerging functionalities in this class of materials.

## METHODS

### Synthesis and characterization:

Single crystals of $Cu_{1+x}Mn_{1-y}SiTe_3$ were synthesized by using the melt-growth method. The detailed procedure is described elsewhere [9]. The actual composition of the single crystals is determined using the EDS. The magnetic measurements were carried out in a vibrating-sample SQUID magnetometer (Quantum Design, SQUID-VSM). Heat capacity, four-probe resistivity, and magnetoresistance were measured in a Physical Property Measurement System (PPMS, Quantum Design). Seebeck coefficient was measured using the thermal transport option (TTO) in PPMS. The SHG characterization of single crystals was carried out in a normal-incidence reflection geometry using an 800 nm fundamental laser beam (Solstice Ace, 80 MHz). For SHG polarimetry, the SHG intensity was recorded as a function of the incident polarization angle of linearly polarized light. The resulting polar plots were fitted within the phase symmetry constraints of point-group *m* using the open-source SHAARP package to simulate the SHG response of anisotropic crystals. The linear optical properties were characterized by spectroscopic ellipsometry (Wollam M2000) over the wavelength range of 200-1600 nm. The crystals were mounted with the crystallographic *c*-axis oriented orthogonal to the sample surface. Measurements were performed at multiple angles of incidence (45°-60° relative to the surface normal), and the real and imaginary parts of the refractive index ($n$ and $k$) were obtained by simultaneously fitting the experimental data across all angles. A Lorentz oscillator model was employed to fit the spectra and extract the anisotropic refractive index dispersion (ordinary $n_o$ and extraordinary $n_e$). The lamella for TEM and STEM analysis was prepared using focused ion beam (FIB) techniques. A Thermo Fisher Scientific Helios 660 NanoLab dual-beam FIB was used to thin the sample to electron transparency. TEM images and SAED patterns were acquired using a Thermo Fisher Scientific Talos F200X operated at 200 kV. STEM imaging was conducted at 300 kV using an aberration-corrected Thermo Fisher Scientific Titan G2 microscope. The acquired ADF-STEM image was drift-corrected using MATLAB code developed by Ophus et al. [31].

### Single-crystal neutron diffraction

Single-crystal neutron diffraction experiments were carried out on the TOPAZ diffractometer at Oak Ridge National Laboratory, USA. A single crystal weighing approximately 20 mg was chosen for comprehensive data collection at 300 K, 100 K, and 5 K. A Cryogenic goniometer was adapted to cover the temperature region from 5 K to 300 K. The crystal was rotated through a wide angular range to survey elastic peaks across an extensive portion of reciprocal space. Moreover, the temperature dependence of selected nuclear and magnetic peaks was monitored during data collection as the sample temperature was slowly ramped at 0.6 K/min through $T_N$. The neutron data were processed using Mantid [32], and the integrated Bragg peak intensities were extracted using TOPAZ-specific Python scripts [33]. Finally, the crystal structure was determined utilizing the crystallographic computing system JANA.

**Computation Method:**

The PBE-level [29] spin-polarized density functional theory (DFT) calculation is done by VASP [34–36] with a cutoff energy of 500 eV and a 7×4×7 Monkhorst-Pack k-space sampling mesh [37] in self-consistent calculations with a $1\times10^{-7}$ eV electronic convergence criteria. The residue force after structural relaxation is <0.01 eV/Å for all atoms. The rotationally invariant DFT + U correction [38] is included in self-consistent and band calculations before and after the structural relaxation, with the effective Hubbard U value 3.9 eV for Mn and 7.2 eV for Cu to make the results comparable with previous work [9]. The k-path for the band structure calculation is determined using pymatgen [39,40]. The magnetization is anti-parallelly given for the two Mn atoms in the structure to be consistent with the antiferromagnetism in the sample.

**ACKNOWLEDGMENT**

Major efforts in sample synthesis, characterization, and band-structure calculations were supported by the Two-Dimensional Crystal Consortium–Materials Innovation Platform (2DCC-MIP) under NSF Cooperative Agreement No. DMR-2039351. Structural studies, including single-crystal X-ray diffraction and transmission electron microscopy, as well as partial synthesis efforts, and data analysis, were supported by the U.S. Department of Energy under Grant No. DE-SC0024943. Neutron diffraction research used resources at the Spallation Neutron Source, a DOE Office of Science User Facility operated by the Oak Ridge National Laboratory. The beam time was allocated to TOPAZ on proposal number IPTS-34391.1. SS, SVGA, and VG acknowledge the support from the NSF through the Pennsylvania State University Materials Research Science and Engineering Center (MRSEC) DMR-2011839 (2020-2026).

## Supplementary Information

### of

**Defect Control via Cu Enrichment Enhances Multifunctional Properties in the Polar Semiconductor $Cu_{1+x}Mn_{1-y}SiTe_3$**

### Note S1: Nonlinear coefficient and Second-harmonic generation (SHG)

In second-harmonic generation (SHG), the incident electric field $E^{\omega}$ at frequency $\omega$ induces a nonlinear polarization $P^{2\omega}$ at frequency $2\omega$ within a noncentrosymmetric medium. This interaction is described by the relation ($P_i^{2\omega} = d_{ijk}E_j^{\omega}E_k^{\omega}$) where the indices (*i, j, k*) correspond to the Cartesian polarization components, and $d_{ijk}$ represents the elements of the second-order nonlinear susceptibility tensor. SHG measurements on the Cu-enriched single crystals were carried out using a fundamental excitation at 800 nm, producing a second-harmonic output at 400 nm. The SHG tensor for the point group *m* is represented in Voigt notation. The expression is given by,

$$\begin{pmatrix} d_{11} & d_{12} & d_{13} & 0 & d_{15} & 0 \\ 0 & 0 & 0 & d_{24} & 0 & d_{26} \\ d_{31} & d_{32} & d_{33} & 0 & d_{35} & 0 \end{pmatrix} \quad (1)$$

Here, $d_{11}, d_{12}, d_{13}, d_{15}, d_{24}, d_{26}, d_{31}, d_{32}, d_{33}$ and $d_{35}$ are non-zero SHG coefficients. At the experimental geometry (sample normal || crystallographic c axis) and analyzer ||X and analyzer||Y is given by,

$$I_x^{2\omega} = f_{11}d_{11}\cos(2\theta) + f_{22}d_{22}\sin(2\theta) \quad (2)$$

$$I_Y^{2\omega} = f_{33}d_{26}\cos(\theta)\sin(\theta) \quad (3)$$

Here, $f_{11}, f_{22}$ and $f_{33}$ are the Fresnel coefficients of the sample. Normal-incidence reflection measurements under the two experimental geometries were fitted using the equations. The excellent agreement between experiment and simulation confirms that the Cu-enriched crystal is a single-domain specimen with monoclinic *m* point-group symmetry. From the measurement geometry, we determined the ratio of nonlinear coefficients to be $\frac{d_{11}}{d_{12}} = -0.55$.

### Note S2: Band structure of Cu enriched system

Density functional theory calculations are used to better understand the electronic structure of Cu-enriched samples. In the computational structure model, the major elements (vacancy treated as a type of element) are used for the partial occupancy sites to capture the properties of the major phase in the material. Therefore, when compared to the major phase structure (with a similar treatment on the partial occupancy sites) of the Cu-deficient sample [1], other than a slight change (<0.2%) in the lattice constants, this Cu-enriched structure essentially replaces two Mn atoms with two Cu atoms in the Si-dimer layer. The Cu-enriched model has the Fermi level in the valence band, consistent with the *p*-doped indirect-gapped semiconducting feature observed in the experiments. This p-doping is understandable by counting the electrons in the system; two $Cu^{1+}$ atoms replacing two $Mn^{2+}$ atoms effectively leaves the system losing two electrons, while the rigid band approximation is appropriate here, probably due to the similar orbital interactions between 3d atoms with the surrounding Te atoms in the Si-dimer layer where the replaced Cu atoms do not relax too far away from the octahedron center (**Fig. S1**).

### Note S3: Magnetic relaxation measurements

The Cu-deficient system demonstrated the coexistence of spin glass behavior with long-range AFM magnetic ordering below $T_N$ and a short-range ordering between $T_N$ and 60 K, which was attributed to the presence of stacking faults in the samples [1]. It is noteworthy that with the Cu doping, the stacking faults in the Cu-enriched system have been significantly suppressed. However, we performed magnetic relaxation measurements of the Cu-enriched sample at different temperatures, as shown in **Fig. S3,** confirming no glassy behavior or short-range order present in the system.

### Note S4: Magnetic entropy change ($\Delta S_m$)

We have calculated the magnetic entropy change ($\Delta S_m$) from the measured isothermal *M-H* curves, using the following integrated Maxwell equation [2]:

$$\Delta S_m(T, \Delta H) = \int_0^H \frac{\partial M}{\partial T} dH$$

where *M*, *T*, and *H* are the magnetization of the sample, the instantaneous temperature, and the applied magnetic field, respectively. The metamagnetic transition for *H // b* does not involve any magnetic field-induced hysteresis, implying a second-order type spin-flop transition. The temperature dependence of $\Delta S_m$ as a function of temperature for different fields changes along the *b*-axis, as shown in **Fig. S4A**. As can be seen, $\Delta S_m$ started to show a sign reversal when the field is 4 T or higher. The sign reversal of $\Delta S_m$ is clearer when $\Delta S_m$ is plotted as a function of field at different temperatures in **Fig. S4B**. Interestingly, $\Delta S_m$ does not show any sign changes at higher temperatures where the metamagnetic spin-flop transition vanishes. For the *H // b* condition, a maximum $\Delta S_m$ of ~ 0.07 J $kg^{-1}K^{-1}$ is observed at 10 K in the presence of a 7 T magnetic field.

We further calculated the temperature dependence of $\Delta S_m$ for different field changes while *H // c*, where we did not observe any metamagnetic behavior (**Figure S5A**), and it is plotted in **Figure S5B**. For the *H // c* condition, the peak value of $\Delta S_m$ is ~ 0.09 J $kg^{-1}K^{-1}$ K in the presence of a 7 T magnetic field. As can be seen, $\Delta S_m$ does not demonstrate any sign reversal feature, confirming that the metamagnetic spin flop transition is associated with the sign reversal of $\Delta S_m$.

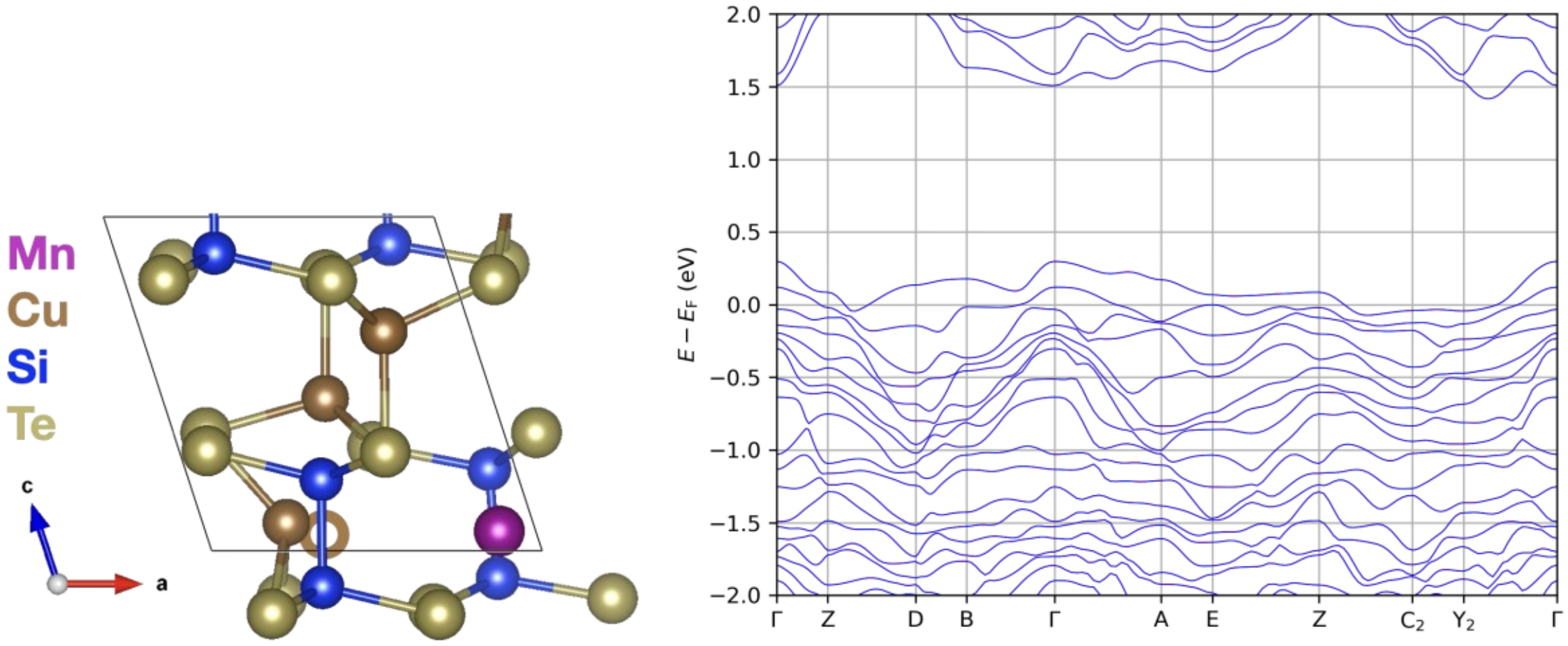

**Figure S1:** The visualization of the relaxed structure model of $Cu_6Mn_2Si_4Te_{12}$ (left). The most significant atomic position change is from the Cu in the Si-dimer layer relaxing to the (–a)-direction. The open brown circle shows the Te-octahedron center site of the Cu before relaxation. The band structure before structural relaxation (right). The band structure is spin-degenerate. The direct/indirect band gaps are 1.21/1.12 eV.

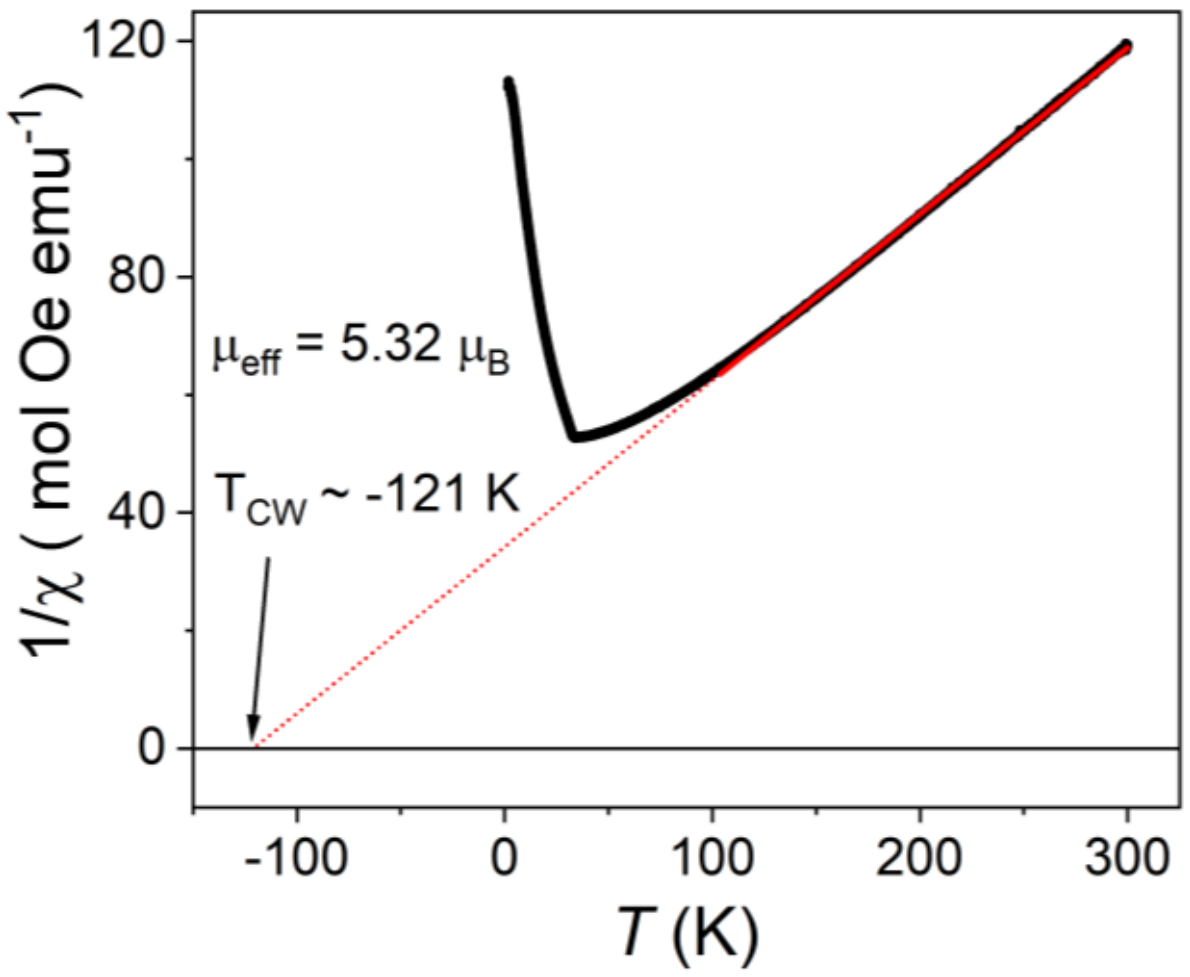

**Figure S2**: The inverse dc magnetic susceptibility of S#2 as a function of temperature under $H$ = 1 kOe applied along the polar $b$ axis. High-temperature data is well-fitted by the Curie-Weiss law, as shown by the red solid line. The negative Curie-Weiss temperature is extracted to be $T_{CW}$ = -121 K, which suggests that the magnetic order below $T_N$ is of AFM character.

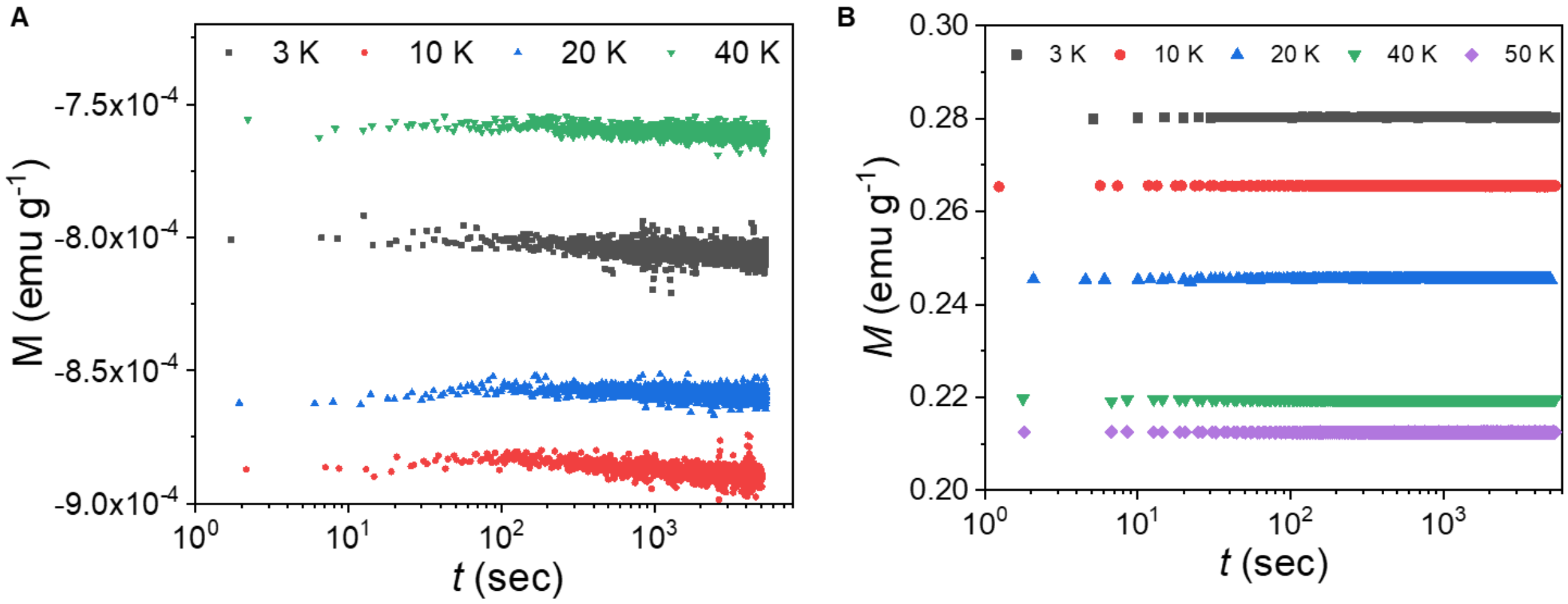

**Figure S3**: Magnetic relaxation measurements of sample S#2, at various temperatures. We followed two measurement procedures: A) The sample was cooled to the target temperature, the magnetic field was increased to 0.5 T, and subsequently reduced to 0 T, and after that, magnetization was measured as a function of time. The sample itself shows low magnetization, and the sample holder dominates the measured signal, resulting in a negative magnetization. (B) The sample was cooled to the target temperature, the magnetic field was increased to 0.5 T, and magnetization was measured as a function of time in the presence of the magnetic field. In both cases, the initial magnetization is almost insensitive to time, indicating that no glassy behavior or short-range ordering is present.

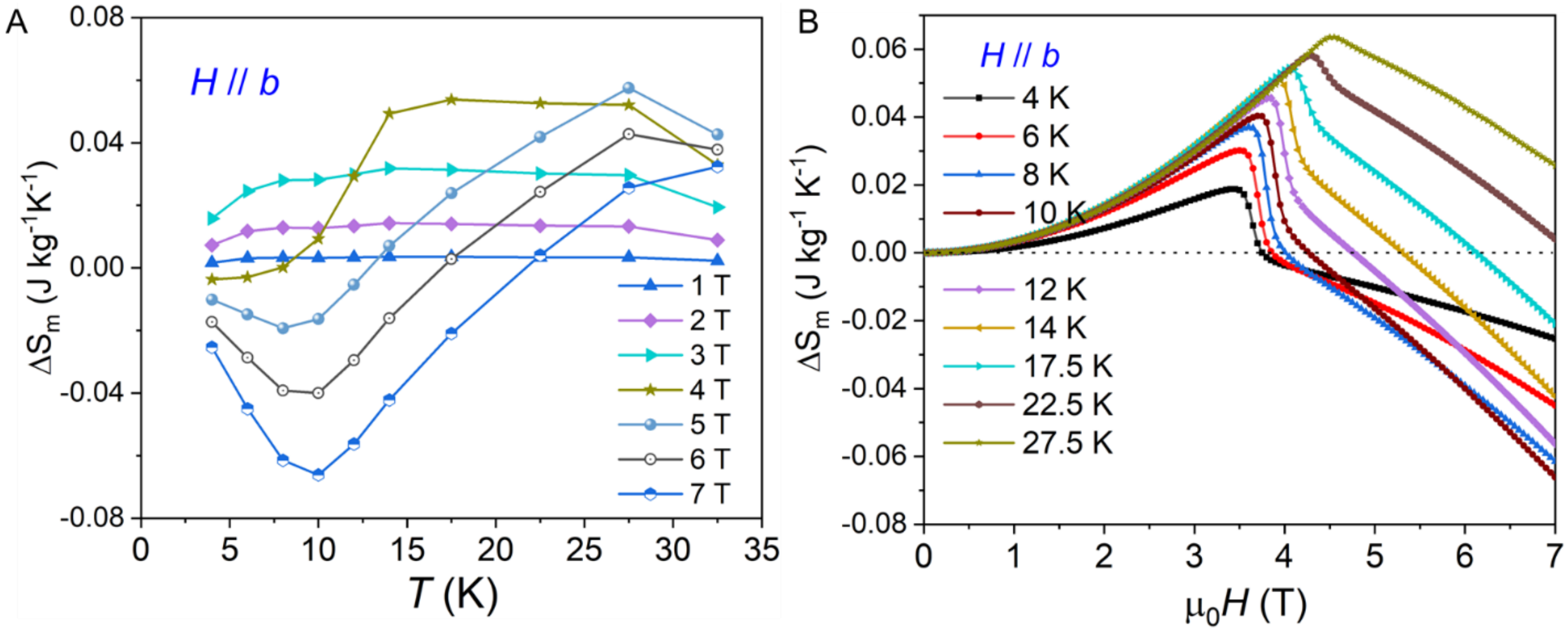

**Figure S4**: (A) Magnetic entropy change ($\Delta S_m$) of sample S#2 as a function of temperature for various magnetic fields along the polar *b*-axis. (B) $\Delta S_m$ plotted as a function of applied magnetic field at different temperatures.

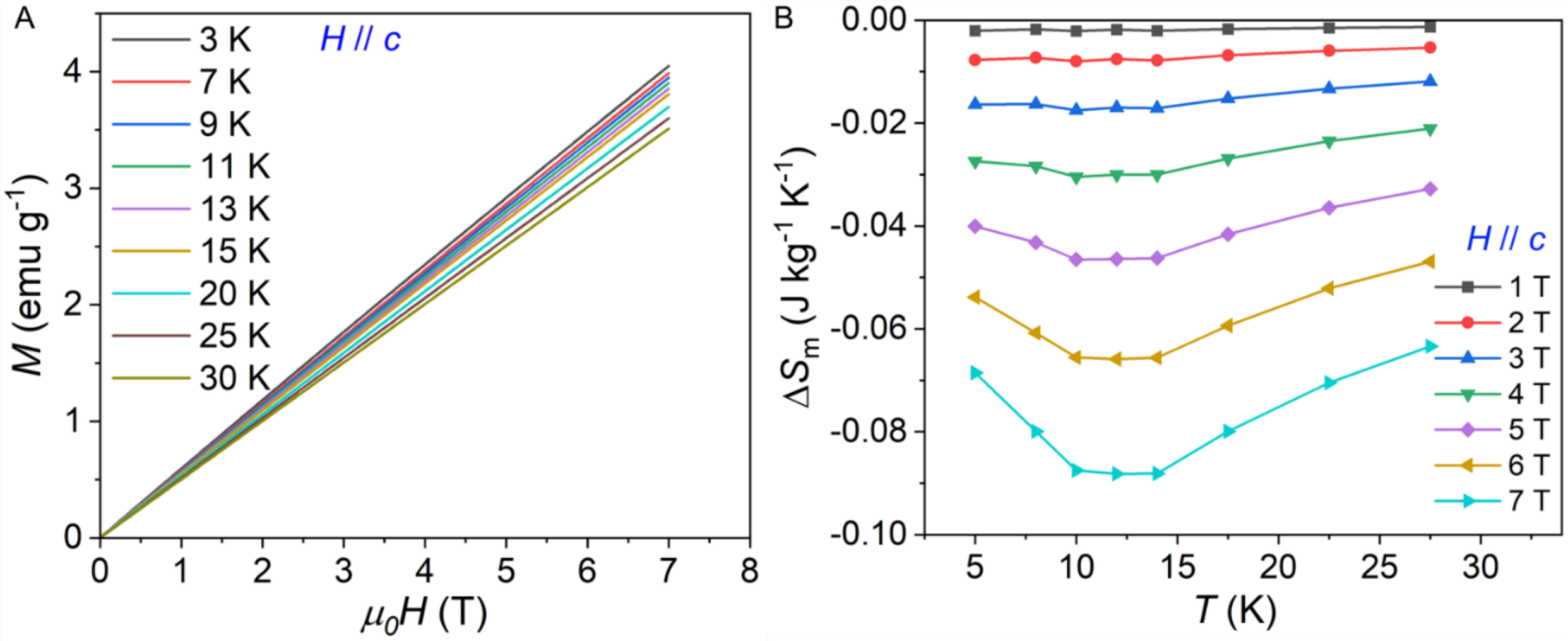

**Figure S5**: (A) Isothermal *M-H* of S#2 at various temperatures below $T_N$. (B) $\Delta S_m$ as a function of temperature for different applied magnetic fields along the polar b axis.

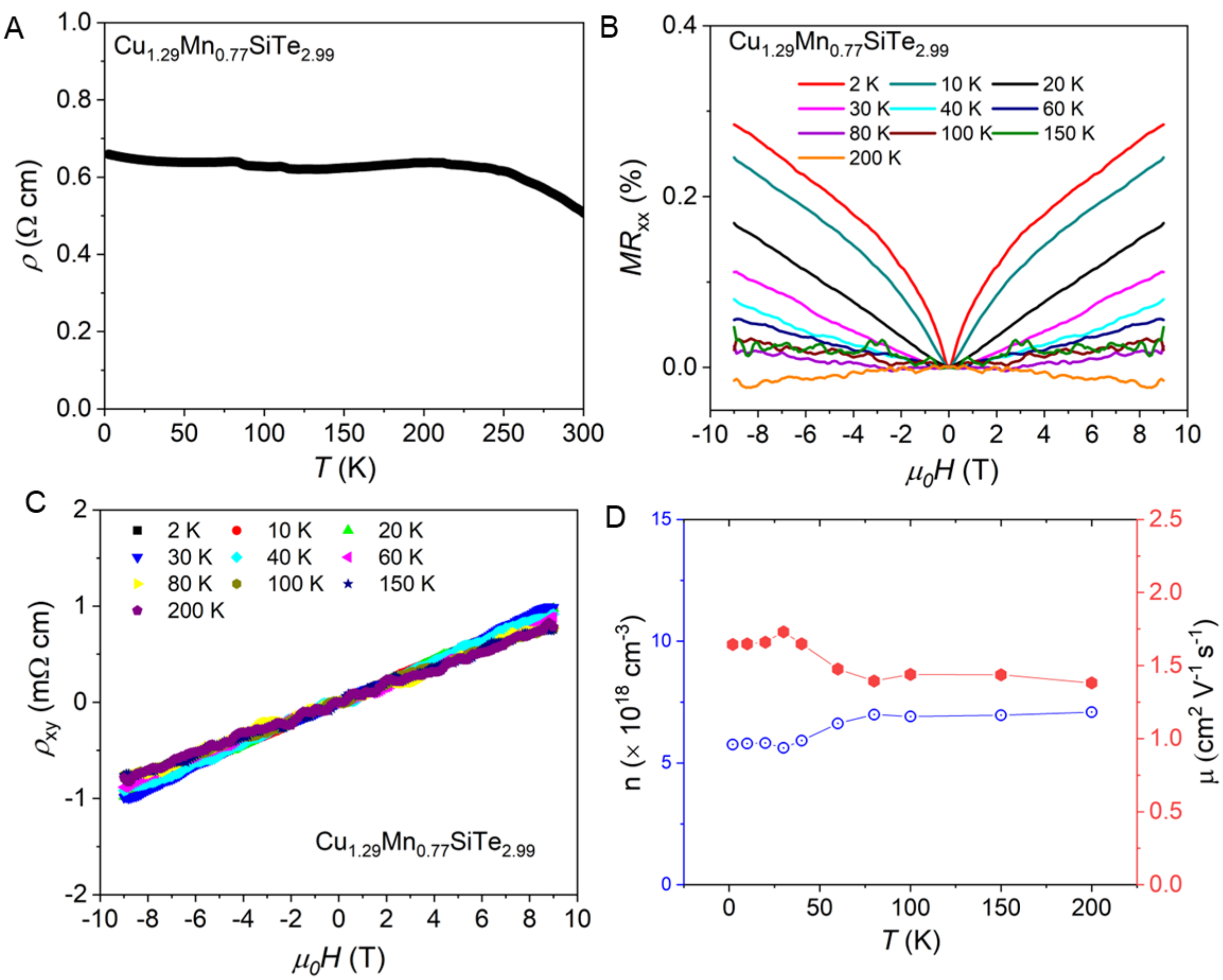

**Figure S6**: **Magnetotransport properties of $Cu_{1.29}Mn_{0.77}SiTe_{2.99}$ (S#5).** (A) Temperature dependence of zero-field longitudinal resistivity ($\rho_{xx}$). The magnetic field dependence of in-plane (B) magnetoresistance and (C) hall resistivity of $Cu_{1.16}Mn_{0.73}SiTe_{2.91}$ (S#2) at various temperatures. (D) The calculated carrier density and carrier mobility as a function of temperature for the S#5 sample.

**Table S1:** XRD refinement parameters.

| **Temperature (K)** | 293(2) |
|---|---|
| **Pressure (GPa)** | Ambient |
| **Space Group** | $P2_1/m$ |
| **Unit Cell dimensions** | $a$ = 7.0925(14) Å<br>$b$ = 12.304(3) Å<br>$c$ = 7.5388(15) Å<br>$\beta$ = 107.94(3)° |
| **Volume** | 625.9(2) $Å^3$ |
| **Z** | 4 |
| **Density (calculated)** | 5.466 |
| **Absorption coefficient** | 18.765 |
| **F (000)** | 872.0 |
| **2θ range** | 7.654 to 82.492 |
| **Reflections collected** | 4260 |
| **Independent reflections** | 4260 [$R_{int}$ = 0.0657] |
| **Data/restraints/parameters** | 4260/0/77 |
| **Final R indices** | $R_1$ (I>2σ(I)) = 0.0413; w$R_2$ (I > 2σ(I)) = 0. 1166<br>$R_1$ (all) = 0.0450; w$R_2$ (all) = 0.1187 |
| **Largest diff. peak and hole** | +4.12 $e^-/Å^3$ and -3.82 $e^-/Å^3$ |
| **Goodness-of-fit on $F^2$** | 1.140 |

**Table S2:** The following samples with actual compositions from EDS are used in this study to investigate optical, magnetic, and transport properties.

| Sample name | EDS Composition |
|---|---|
| S#1 | $Cu_{1.14}Mn_{0.77}SiTe_{2.89}$ |
| S#2 | $Cu_{1.10}Mn_{0.80}SiTe_{2.80}$ |
| S#3 | $Cu_{1.14}Mn_{0.78}SiTe_3$ |
| S#4 | $Cu_{1.16}Mn_{0.73}SiTe_{2.91}$ |
| S#5 | $Cu_{1.29}Mn_{0.77}SiTe_{2.99}$ |
| S#6 | $Cu_{1.24}Mn_{0.91}SiTe_{3.03}$ |